\documentclass[prc,twocolumn,aps,showpacs,preprintnumbers, tightenlines,showpacs,floatfix,superscriptaddress,nofootinbib]{revtex4-1}

\usepackage{bm}
\usepackage{soul}
\usepackage{color}
\usepackage{hyperref}
\usepackage{amsmath}
\usepackage{graphicx,epsfig}
\usepackage{longtable}
\usepackage{xcolor}
\usepackage{color}
\usepackage{float}

\usepackage{dcolumn}     %
\usepackage[resetlabels,labeled]{multibib}
\newcites{Math}{Mathgg}

\def\beq{\begin{equation}}
\def\eeq{\end{equation}}
\def\beqy{\begin{eqnarray}}
\def\eeqy{\end{eqnarray}}

\newcommand{\pkuphy}{School of Physics, Peking University, Beijing 100871,
China}
\newcommand{\gscaep}{Graduate School of China Academy of Engineering Physics, Beijing 100193, China}
\newcommand{\chep}{Center for High Energy Physics, Peking University, Beijing 100871, China}
\newcommand{\ccqm}{Collaborative Innovation Center of Quantum Matter, Beijing 100871, China}

\begin{document}

{

\title{Multi-reference Trial State for Lattice Quantum Monte Carlo Simulations}

\author{Teng~Wang}\email{tenggeer@pku.edu.cn}\affiliation{\pkuphy}
\author{Xu~Feng}\email{xu.feng@pku.edu.cn}\affiliation{\pkuphy}\affiliation{\chep}\affiliation{\ccqm}
\author{Bing-Nan~Lu}\email{bnlv@gscaep.ac.cn}\affiliation{\gscaep}

\begin{abstract}
Nuclear lattice effective field theory (NLEFT) is an efficient \textit{ab initio} tool for solving nuclear many-body systems using the imaginary-time projection technique, where the preparation of trial states is essential for substantially reducing the computational cost required to achieve the desired numerical precision. 
It has been challenging in forming optimal multi-reference trial states using multiple Slater determinants within auxiliary-field based quantum Monte Carlo frameworks like NLEFT.
In this work, we develop a novel sampling method for efficiently incorporating such multi-reference trial states into NLEFT calculations. We applied the optimized trial state to $^7$Li and $^8$Li, finding overall improvements in calculated energies, electromagnetic properties, and transitions compared to results obtained without these optimizations.
Our approach provides a reliable foundation for accurately simulating nuclear ground and low-lying excited states within the NLEFT framework.

\end{abstract}

\maketitle
}

\section {Introduction}
\label{sec:intro}

Nuclear Lattice Effective Field Theory (NLEFT) is an efficient \textit{ab initio} method for solving nuclear many-body problems~\cite{Lee:2009,Lahde:2019, Dean:2025}. It employs effective field theory (EFT) principles and treats protons and neutrons as fundamental degrees of freedom, with the nuclear force implemented on a cubic lattice. Starting from trial states possessing the prescribed quantum numbers, the ground state and low-lying excited states are extracted through Euclidean-time projection calculations performed using the quantum Monte Carlo (QMC) technique. With the rapid development of advanced algorithms, NLEFT has become a powerful tool for investigating a wide range of nuclear properties. These include nuclear binding energies~\cite{Elhatisari:2022, Song:2025, EPJA31-105, PhysRevLett.104.142501, EPJA45-335, PLB732-110}, charge radii~\cite{Ren:2025,Konig:2023}, nuclear electromagnetic properties~\cite{Shen:2023,Shen:2025}, $\beta$-decay lifetimes~\cite{Elhatisari:2024gg, Teng:2025_decay}, hypernuclei~\cite{EPJA56-24,Hildenbrand:2022, EPJA60-215}, hyper-neutron matter~\cite{Tong:2024, Tong:2025}, and hot nuclear matter~\cite{Ma:2023, Ren:2023, Lu:2020}.

Generally, under the action of the imaginary time evolution operator, any trial state possessing non-zero overlap will ultimately evolve to the true ground state for sufficiently large projection time. In realistic calculations, however, the Monte Carlo sampling encounters the notorious sign problem~\cite{PRL94-170201}, stemming from the anti-symmetrization of fermionic wave functions. This sign problem leads to an exponentially suppressed signal-to-noise ratio with increasing particle number, volume, and projection time. Consequently, the application of nuclear QMC is severely constrained by the rapid growth of statistical errors. In many important cases, calculations are restricted to small imaginary time where the sign problem remains manageable, while extrapolations to infinite projection time become highly challenging~\cite{Lahde:2014,Lahde:2015}. 

In recent years, substantial efforts have been devoted to mitigating the sign problem in NLEFT, including the development of sign-problem-free leading-order  interactions~\cite{Lu:2019, Niu:2025}, the application of second-order perturbation theory~\cite{Lu:2021, Liu:2025}, the development of the wavefunction matching method~\cite{Elhatisari:2022}, and the rank-one operator method~\cite{Ma:2023}. These methods have proven efficient at reducing the sign problem in many important scenarios, particularly for light even-even nuclei, where low-lying states are well separated by significant energy gaps and their interferences diminish rapidly during imaginary time projection. However, most nuclei possess odd numbers of protons or neutrons, resulting in much denser low-lying states with reduced energy gaps above the true ground state. Furthermore, important observables such as charge radii converge much more slowly than binding energies due to crossing terms between the ground state and excited states~\cite{Ren:2025}. In such cases, significant excited-state contamination may persist before the signal is overwhelmed by noise, leading to substantial extrapolation errors in energies and other observables.

A direct approach to accelerating convergence involves optimizing the trial state. Constructing a sufficiently optimal trial state that mimics the true ground state suppresses excited-state contamination from the outset and significantly accelerates convergence. 
Improving trial states using variational principles in nuclear QMC calculations has a long history, specifically in variational Monte Carlo calculations~\cite{Pandharipande:1979}. In such studies, Jastrow-like operators capture short-range nucleon correlations, while linear combinations of different shell-model Slater determinants reproduce long-range correlations. These optimized wave functions can then be used in Green's function Monte Carlo and auxiliary field diffusion Monte Carlo~\cite{Carlson:1987, Carlson:1988,Schmidt:1999,QMCreview} calculations to achieve rapid convergence to the infinite imaginary time limit. Recently, artificial neural networks have also been employed to emulate complex many-body correlations in trial wave functions for variational QMC~\cite{RMP94-031003}. 
Combining multiple Slater determinants is a common technique in beyond-mean-field calculations to incorporate collective many-body correlations into the wave functions~\cite{Bender:2003, Meng:2005}. 
For instance, nuclear shape oscillations can be described by superposing a series of Slater determinants with different deformations, a method known as the generator coordinate method~\cite{Reinhard:1987}. More complex quantum correlations can be built upon such multi-reference wave functions.

While the multi-reference techniques are quite successful in mean-field models and shell-model based \textit{ab initio} methods~\cite{Hergert:2016}, most NLEFT calculations limit the trial wave functions to a single Slater determinant. The challenge arises because NLEFT employs the auxiliary-field quantum Monte Carlo (AFQMC) technique within the second-quantized representation, which decouples two- and three-body nuclear forces into sums of one-body operators interacting with background auxiliary fields.
In AFQMC, single-particle wave functions interact independently with these auxiliary fields, effectively reducing the problem to a solvable one-body system. Many-body correlations emerge only upon integrating out the auxiliary fields using the Monte Carlo algorithm. 
Consequently, it is numerically simplest and most economical to define the trial state as a single Slater determinant, formed from an antisymmetrized product of single-particle wave functions. 
Incorporating many-body correlations directly into the trial state necessitates multi-channel sampling of the auxiliary fields and substantially increases computational demands. For this reason, although previous NLEFT studies have utilized a plethora of different trial states, such as shell-model-like states~\cite{Lu:2019, Niu:2025, Hildenbrand:2025}, $\alpha$-cluster states~\cite{Epelbaum:2009a,PhysRevLett.109.252501, Lu:2019, Shen:2021, Shen:2023}, and plane-wave states~\cite{Elhatisari:2017}, multi-reference trial states have rarely been applied. 
The primary exceptions involve projecting onto definite total momentum and angular momentum (cubic group representations) by summing over positions and orientations of localized wave functions~\cite{Epelbaum:2009a,PhysRevLett.109.252501, Shen:2023}, as well as the coupled channel calculation of  the alpha-particle monopole transition form factor~\cite{Ulf:2024}.

Recent years have witnessed significant improvements in the numerical precision of NLEFT calculations. To further reduce the extrapolation errors, systematically incorporating multi-reference trial wave functions into the QMC framework is highly desirable. In this work, we firstly discuss the application of multi-reference trial state based on shell model configurations in NLEFT. We construct the trial state from shell-model single particle wave functions based on a deformed harmonic oscillator potential. We optimize the oscillator frequency and deformation parameters, then calculate correlation functions between Slater determinants generated from different valence nucleon configurations. Subsequently, we developed an efficient algorithm for sampling and calculating these correlation functions within a shared Monte Carlo ensemble. Testing this method on $^{7}$Li and $^{8}$Li reveals that the optimized trial state exhibits significantly reduced excited-state contamination. Accelerated convergence is simultaneously observed for both the binding energy and electromagnetic properties, enabling robust and accurate imaginary-time extrapolations.

This paper is organized as follows: Section~\ref{sec:lat interaction} describes the nuclear interaction used in this work; Section~\ref{sec:cal method} discusses the trial state optimization method and relevant technical details; Section~\ref{sec:Res} presents the improvement results for $^7$Li and $^8$Li. Additional details are provided in the Appendix.

\section{Interaction And Lattice Setup}
\label{sec:lat interaction}

In this work, we use the interaction recently developed in Ref,~\cite{Niu:2025}, with lattice spacing $a=1.32$ fm. The interaction comprises a
two-body and a three-body contact term, which are both SU4-invariant and smeared locally and nonlocally. An additional spin-orbit
coupling term is supplemented to fine-tune the shell structure. This simple interaction achieves precise predictions of binding
energies for 76 even-even nuclei, matching state-of-the-art phenomenological mean-field models. The binding energies of $^{100}$Sn and $^{132}$Sn are also calculated, whose relative errors against the experimental data are below 1$\text{\textperthousand}$ and 3$\%$, respectively. The unique advantage of this interaction is that it is rigorously free from sign problems for even-even nuclei, which allows an accurate non-perturbative solution of the Hamiltonian. For the odd-even nucleus $^{7}$Li and odd-odd nucleus $^{8}$Li studied here, we find that the sign problem is also under good control, so the system can evolve for a sufficiently long time. This helps us better identify the convergence rate of different trial states to validate the advantage of our method. All the following calculations are performed on an $L=14.52$ fm cubic lattice without quantifying finite-volume uncertainties, which is of little relevance to the purpose of this work.

\section {Method}
\label{sec:cal method}

\subsection{The construction and improvement of trial states}
\label{Trial State Improvement}

The building block of our trial state is the solution of the deformed harmonic oscillator,
\begin{equation}
\label{H_ho}
    H_{\mathrm{ho}} = \frac{\boldsymbol{p}^2}{2m} + \frac{1}{2}m\omega^2\left[\left(1-\frac{4}{3}\delta\right)z^2+\left(1+\frac{2}{3}\delta\right)(x^2+y^2)\right],
\end{equation}
with $m=938.92$ MeV the nucleon mass, $\omega$ the oscillator frequency and $\delta$ the deformation parameter. The solution of $H_{\mathrm{ho}}$ is the mean-field orbit $|\phi_n(\omega,\delta)\rangle$, where $n$ denotes the set of quantum numbers of the orbit. By filling $A$ nucleons into these orbits, one can construct a $A$-body Slater determinant,
\begin{equation} |\Phi^{\lambda}_{J_z, P}(\omega,\delta)\rangle = |\phi_{\lambda(1)}(\omega,\delta)\rangle\wedge\cdots\wedge|\phi_{\lambda(A)}(\omega,\delta)\rangle,
\end{equation}
where $J_z$ is the total magnetic quantum number and $P$ is the parity. The indices $\lambda\equiv \{\lambda(1),\cdots,\lambda(A)\}$ label the orbits building the Slater determinant.  Since the single Slater determinant is generally not the eigenstate of $J^2$ even if $\delta=0$ and $H_{\mathrm{ho}}$ is rotational invariant, it cannot uniquely determine the total angular momentum of the system.  To address the issue, we consider $N_d$ different shell distributions labeled $d_1, d_2,\cdots, d_{N_d} $, which are distinct from each other by the occupation number of each shell. For a given distribution $d_i$, we construct the following trial state, 
\begin{equation}
\label{Psi_conf}  |\Psi^{d_i}_{J_z=J,P}(\omega,\delta)\rangle = \sum_{\lambda\in d_i}a^{d_i}_{\lambda}|\Phi_{J_z=J, P}^{\lambda}(\omega,\delta)\rangle.
\end{equation}
In the above definition, the summation runs over Slater determinants belonging to $d_i$.
The coefficients $a^{d_i}_\lambda$ are fixed by the requirement that $|\Psi^{d_i}_{J_z=J,P}\rangle$ carries the total angular momentum $J$ for $\delta=0$, and the magnetic quantum number $J_z$ is set as $J$.

\begin{figure*}
\begin{center}  \includegraphics[height=2.9in]{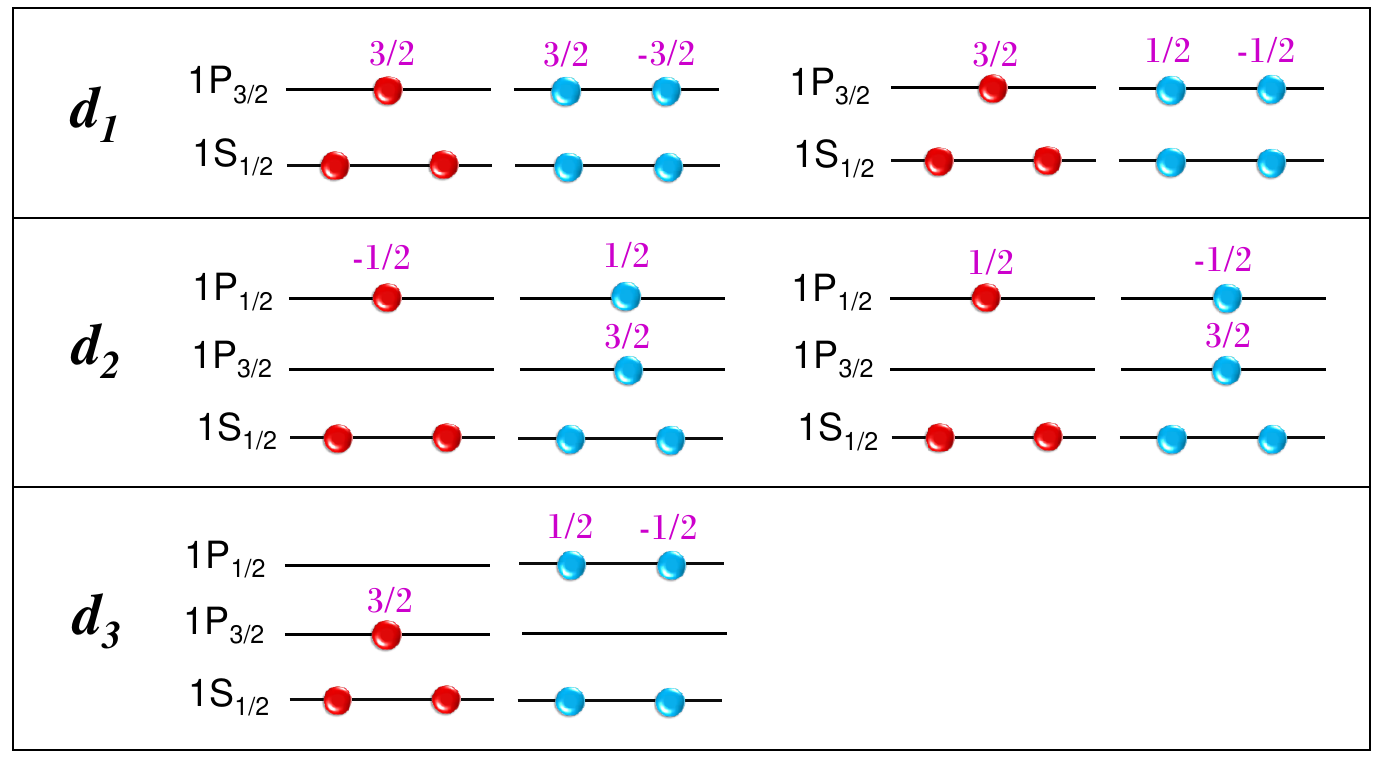}
\caption{The three different shell distributions, $d_1$, $d_2$ and $d_3$, for constructing the trial state of $^7$Li(3/2$^-$). The red (blue) circle represents the occupied orbit of the proton (neutron). The number close to the circle denotes the magnetic quantum number of the $p$-shell orbit. $d_1$ consists of two Slater determinants shown in the first row. $d_2$ and $d_3$ comprises two and one Slater determinants, respectively.  }
\label{fig:Li7_conf} 
\end{center}
\end{figure*}

As an illustration, in Figure~\ref{fig:Li7_conf}, we show all the three distributions used for the calculation of $^7$Li(3/2$^-$), denoted as $d_1$, $d_2$ and $d_3$. Slater determinants subject to a common distribution are plotted in the same row, which are all eigenstates of $J_z$ with eigenvalue $3/2$.  They are combined into 3 trial states, $|\Psi^{d_1}_{J_z=3/2,-}\rangle, |\Psi^{d_2}_{J_z=3/2,-}\rangle$ and $|\Psi^{d_3}_{J_z=3/2,-}\rangle$, which carry $J=3/2$ when $\delta=0$.  In the appendix, we also show the trial states for $^{7}$Li(1/2$^-$), $^{8}$Li(2$^+$) and $^{8}$Li(1$^+$). 

The above discussions lay the foundation for the systematic improvement of trial states, which consists of two steps:

\textbf{Step 1}. We first optimize the value of $\omega$ and $\delta$. Specifically, the shell-model ground-state configuration, such as $d_1$ in Figure~\ref{fig:Li7_conf}, is used to calculate selected observables. The evolution of these observables versus the imaginary time is then plotted, as will be shown in the next section. We look for the optimized values, $\omega=\omega^*$ and $\delta=\delta^*$, for which the overall convergence is the fastest. They are fixed for the next step of optimization. 

\textbf{Step 2}.  In the second step, we incorporate the effect of particle-hole excitations. To achieve this, we consider the linear combination of the $N_d$ distributions in Eq.~(\ref{Psi_conf}),
\begin{equation}
\label{Psi_opt}|\Psi^{\mathrm{opt}}_{J_z=J, P}\rangle = \sum_{i=1}^{N_d}b_i|\Psi^{d_i}_{J_z=J, P}(\omega^*,\delta^*)\rangle,
\end{equation}
where $b_i$ is variationally determined to optimize $|\Psi^{\mathrm{opt}}\rangle$.  Specifically, we calculate the following correlation functions between different Slater determinants,
\begin{eqnarray}  
\label{eq: C_and_O}
&C_{\lambda\lambda'}(\tau)&\equiv\langle\Phi_{J_z=J,P}^{\lambda}|e^{-H\tau}|\Phi_{J_z=J,P}^{\lambda'}\rangle,\nonumber\\
&O_{\lambda\lambda'}(\tau)&\equiv\langle\Phi_{J_z=J,P}^{\lambda}|e^{-H\tau/2}Oe^{-H\tau/2}|\Phi_{J_z=J,P}^{\lambda'}\rangle,
\end{eqnarray}
where $\tau$ is the projection time and $O$ is an arbitrary interpolating operator. They are then grouped into correlation functions between trial states of different shell distributions,
\begin{eqnarray}
\label{correlation_conf}
C^{ij}(\tau)&\equiv&\langle\Psi_{J_z=J,P}^{d_i}|e^{-H\tau}|\Psi_{J_z=J,P}^{d_j}\rangle \nonumber\\
&=&\sum_{\lambda\in d_i}\sum_{\lambda'\in d_j}(a^{d_i}_\lambda)^*  a^{d_j}_{\lambda'} C_{\lambda\lambda'}(\tau),\nonumber\\
O^{ij}(\tau)&\equiv&\langle\Psi_{J_z=J,P}^{d_i}|e^{-H\tau/2}Oe^{-H\tau/2}|\Psi_{J_z=J,P}^{d_j}\rangle\nonumber\\
&=&\sum_{\lambda\in d_i}\sum_{\lambda'\in d_j}(a^{d_i}_\lambda)^*  a^{d_j}_{\lambda'} O_{\lambda\lambda'}(\tau),
\end{eqnarray}
 based on which the expectation value of the Hamiltonian $H$ with respect to $|\Psi^{\mathrm{opt}}\rangle$ can be expressed as follows,
\begin{eqnarray}
\label{E_opt}
E(\tau)&\equiv&\frac{\langle\Psi^{\mathrm{opt}}_{J_z=J, P}|e^{-H\tau/2}He^{-H\tau/2}|\Psi^{\mathrm{opt}}_{J_z=J, P}\rangle}{\langle\Psi^{\mathrm{opt}}_{J_z=J, P}|e^{-H\tau}|\Psi^{\mathrm{opt}}_{J_z=J, P}\rangle}\nonumber\\
&=&\frac{\sum_{i,j=1}^{N_d}b_i^* b_{j} H^{ij}(\tau)}{\sum_{i,j=1}^{N_d}b_i^* b_{j} C^{ij}(\tau)}.
\end{eqnarray}
By minimizing the value of $E(\tau)$, $b_i$ can be solved from the following $N_d\times N_d$ generalized eigenvalue equation,
\begin{equation} 
\label{GEVP_equation}\sum_{j=1}^{N_c}H^{ij}(\tau)b_{j} = E(\tau) \sum_{j=1}^{N_c}C^{ij}(\tau)b_{j}.
\end{equation}
To show the effect of $|\Psi^{\mathrm{opt}}\rangle$, we use it to calculate various observables of the target nucleus and compare it to the result of a trial state without improvement, which is denoted as $|\Psi^{\mathrm{compare}}_{J_z=J, P}(\omega_0, \delta_0)\rangle$. $|\Psi^{\mathrm{compare}}\rangle$ is built from  the shell-model ground-state distribution, such as $d_1$ in Figure~\ref{fig:Li7_conf}. The deformation parameter and the oscillator strength are fixed as $\omega_0=41A^{-1/3}$ MeV and $\delta_0=0$, with $\omega_0$ phenomenologically determined from the mean square radius of spherical nuclei~\cite{Bohr1}.

\subsection{The algorithm to calculate the correlation functions between multi-Slater determinants}

\label{Monte Carlo}

The basis of step 2 in the previous subsection is the calculation of the correlation functions, $C_{\lambda\lambda'}$ and $H_{\lambda\lambda'}$, between different Slater determinants. Let the total number of Slater determinants be $N_S$, then $C_{\lambda\lambda'}$  has $N_S^2$ components. Even if $N_S$ is not too large, say $N_S=5$ in the case of $^7$Li(3/2$^-$), generating $N_S^2$ different correlation functions from a common ensemble would be computationally demanding for the AFQMC method, and we overcome this difficulty through a new algorithm developed in this work.  In the following, we first introduce the notation needed to illustrate the new algorithm, together with the traditional method used by the NLEFT Collaboration to sample correlation functions between multi-Slater determinants. We then present the new algorithm and compare it to the old one to show its superiority.

For AFQMC calculations, the imaginary time projector $e^{-H\tau}$ is first decomposed into the product of $L_t=\tau/a_t$ transfer matrices $M=:e^{-a_tH}:$,
with $a_t$ the temporal lattice spacing and :\ : the normal ordering operator. Applying auxiliary transformation to the transfer matrices, we obtain
\begin{equation}
    e^{-H\tau}=\int \mathcal{D}\xi \ e^{-\sum_{n_t=1}^{L_t}\xi^2(n_t)/2}\prod_{n_t=1}^{L_t}\mathcal{M}[\xi(n_t)],
\end{equation}
where $\mathcal{M}[\xi(n_t)]$ is the transformed transfer matrix and $\xi(n_t)$ is the auxiliary field of the $n_t$th time slice. For the explicit expression of $\mathcal{M}[\xi(n_t)]$, see the supplemental material in Ref.~\cite{Niu:2025}. For convenience, we define the following ket and bra, 
\begin{eqnarray}
\label{Slater_evolve}
&|\Phi^{\lambda,R}_{J_z,P}(\xi)\rangle& = \left(\prod_{n_t=1}^{L_t/2}\mathcal{M}[\xi(n_t)]\right)|\Phi^\lambda_{J_z,P}\rangle,\nonumber  \\
&\langle\Phi^{\lambda,L}_{J_z,P}(\xi)|&= \langle\Phi^\lambda_{J_z,P}|\left(\prod_{n_t=L_t/2+1}^{L_t}\mathcal{M}[\xi(n_t)]\right),
\end{eqnarray}
based on which the auxiliary field representation of $C_{\lambda\lambda'}$ and $O_{\lambda\lambda'}$ can be expressed compactly as
\begin{eqnarray}
\label{auxiliary_rep}
   & C_{\lambda\lambda'}(\tau)& = \int\mathcal{D}\xi\ e^{-\xi^2/2}\langle\Phi^{\lambda,L}_{J_z,P}(\xi)|\Phi^{\lambda',R}_{J_z,P}(\xi)\rangle,\nonumber\\
   & O_{\lambda\lambda'}(\tau)& = \int\mathcal{D}\xi\ e^{-\xi/2}\langle\Phi^{\lambda,L}_{J_z,P}(\xi)|O|\Phi^{\lambda',R}_{J_z,P}(\xi)\rangle,\nonumber\\
\end{eqnarray}
where for simplicity the summation  over $n_t$ in the exponential has been suppressed. Remarkably, since $\mathcal{M}[\xi(n_t)]$ is the exponential of  one-body operators, the operation of $\mathcal{M}[\xi(n_t)]$ on one Slater determinant simply yields another Slater determinant. Therefore, both $|\Phi^{\lambda,R}_{J_z, P}(\xi)\rangle$ and $\langle \Phi^{\lambda, L}_{J_z, P}(\xi)|$ can be expressed as
\begin{eqnarray}
\label{wedge}
&|\Phi^{\lambda,R}_{J_z,P}(\xi)\rangle& = |\phi^R_{\lambda(1)}(\xi)\rangle\wedge\cdots\wedge |\phi^R_{\lambda(A)}(\xi)\rangle,\nonumber\\
&\langle\Phi^{\lambda,L}_{J_z,P}(\xi)|& = \langle \phi^L_{\lambda(1)}(\xi)|\wedge\cdots\wedge\langle \phi^L_{\lambda(A)}(\xi)|,
\end{eqnarray}
where we have defined the following single-particle states:
\begin{eqnarray}
&|\phi_{\lambda(k)}^R(\xi)\rangle&=\left(\prod_{n_t=1}^{L_t/2}\mathcal{M}[\xi(n_t)]\right)|\phi_{\lambda(k)}\rangle,\\
&\langle\phi_{\lambda(k)}^L(\xi)|& = \langle\phi_{\lambda(k)}|\left(\prod_{n_t=L_t/2+1}^{L_t}\mathcal{M}[\xi(n_t)]\right).
\end{eqnarray}

We now review the traditional method for sampling correlation functions between multi-Slater determinants, which is developed by the NLEFT collaboration~\cite{Lahde:2019} and will be denoted as {\ttfamily MultSlat-Sampler\uppercase\expandafter{\romannumeral1}} in the following. It has been applied to the study of low-lying states of $^{12}$C~\cite{PhysRevLett.109.252501, Epelbaum:2009a,  Shen:2021, Shen:2023}and $^{16}$O~\cite{Epelbaum:2013}, as well as other multi-channel calculations. The idea of this method is to treat the Slater-determinant index $\lambda$ as a random variable and to generate Monte Carlo configurations of $\xi$ and $\lambda$ from the following probability distribution,
\begin{equation}    P_{\lambda\lambda'}(\xi)=\frac{1}{Z} e^{-\xi^2/2}|\langle\Psi^{\lambda,L}_{J_z, P}(\xi)|\Psi^{\lambda',R}_{J_z, P}(\xi)\rangle|,
\end{equation}
with $Z$ the normalization factor. Notice that  for any operator $O$, its correlation function $O_{\lambda\lambda'}$ in Eq.~(\ref{eq: C_and_O}) can be rewritten as
\begin{eqnarray}
 &&O_{\lambda\lambda'}(\tau)
  \nonumber\\
&=&\sum_{\eta\eta'}O_{\eta\eta'}(\tau)\delta_{\lambda\eta}\delta_{\lambda'\eta'}\nonumber\\
  &=&Z\left(\sum_{\eta\eta'}\int\mathcal{D}\xi\ P_{\eta\eta'}(\xi)\right)\frac{\langle\Psi^{\eta,L}_{J_z, P}(\xi)|O|\Psi^{\eta',R}_{J_z, P}(\xi)\rangle}{|\langle\Psi^{\eta,L}_{J_z, P}(\xi)|\Psi^{\eta',R}_{J_z, P}(\xi)\rangle|}\delta_{\lambda\eta}\delta_{\lambda'\eta'},\nonumber\\
\end{eqnarray}
so if the number of Monte-Carlo configurations is equal to $N_{\mathrm{MC}}$, and if we label the auxiliary fields and Slater-determinant indices as $\xi_k$ and $(\eta_k,\eta'_{k})$ for the $k$th configuration, respectively, then the ratio between $C_{\alpha\alpha'}$ and $C_{\beta\beta'}$ is given by
\begin{eqnarray}
  &&C_{\alpha\alpha'}(\tau):C_{\beta\beta'}(\tau) \nonumber\\
  &=&\sum_{k=1}^{N_{\mathrm{MC}}}\frac{\langle\Psi^{\eta_k,L}_{J_z, P}(\xi)|\Psi^{\eta'_k,R}_{J_z, P}(\xi)\rangle}{|\langle\Psi^{\eta_k,L}_{J_z, P}(\xi)|\Psi^{\eta'_k,R}_{J_z, P}(\xi)\rangle|}\delta_{\alpha\eta_k}\delta_{\alpha'\eta'_k}\nonumber\\  &:&\sum_{k=1}^{N_{\mathrm{MC}}}\frac{\langle\Psi^{\eta_k,L}_{J_z, P}(\xi)|\Psi^{\eta'_k,R}_{J_z, P}(\xi)\rangle}{|\langle\Psi^{\eta_k,L}_{J_z, P}(\xi)|\Psi^{\eta'_k,R}_{J_z, P}(\xi)\rangle|}\delta_{\beta\eta_k}\delta_{\beta'\eta'_k},
\end{eqnarray}
which is exact in the limit of $N_{\mathrm{MC}}\rightarrow\infty$. Similar equations also hold for the ratio between $C_{\alpha\alpha'}$ and $O_{\beta\beta'}$. Clearly, the $k$th configuration only contributes to a single component of the correlation function, that is,  $C_{\eta_k\eta_k'}$, while the rest $N_S^2-1$ components are not influenced. Therefore, as $N_S$ increases,  this sampling method would become less and less efficient.

\begin{figure*}
\centering
\includegraphics[height=1.6in]{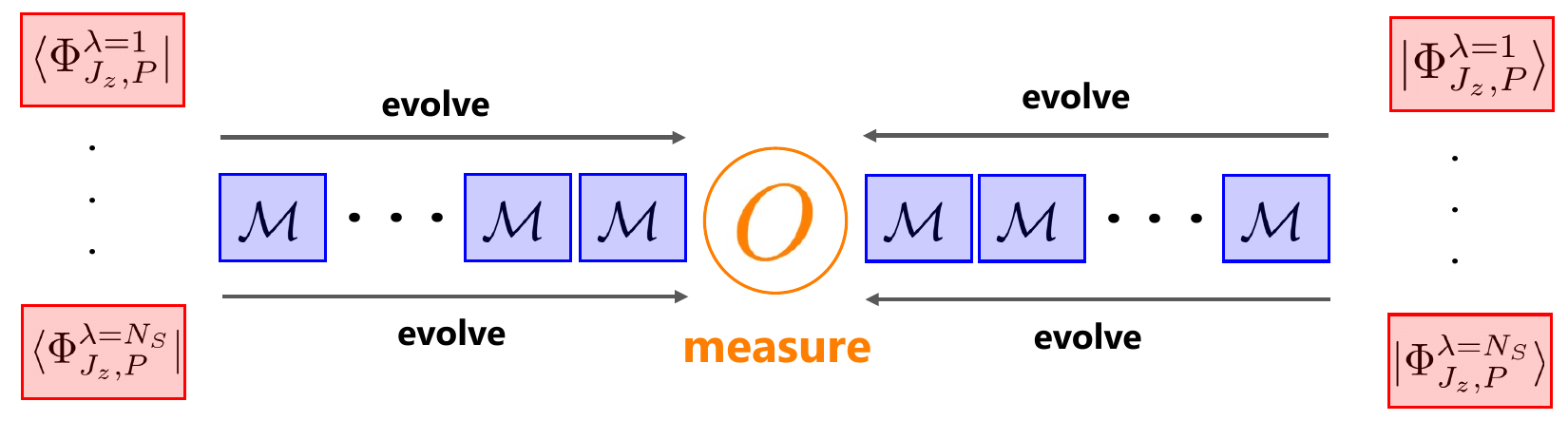}
\caption{The brute-force way to calculate the matrix element $\langle \Phi^{\lambda, L}_{J_z, P}(\xi)|O|\Phi^{\lambda, R}_{J_z, P}(\xi)\rangle$. The $N_S$ Slater determinants are first independently evolved to the middle of the projection time under the operation of the transfer matrices $M$, the $N_S^2$ matrix elements are then calculated individually. }
\label{fig:direct_sample}
\end{figure*}

\begin{figure*}
\centering
\includegraphics[height=1.6in]{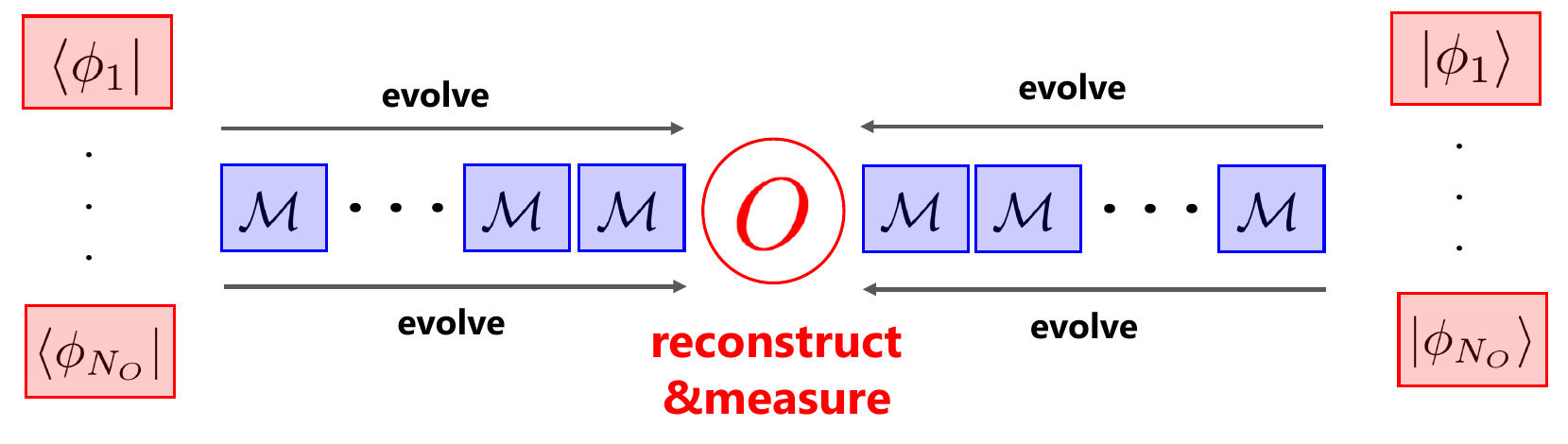}
\caption{The efficient way to calculate the matrix element $\langle \Phi^{\lambda, L}_{J_z, P}(\xi)|O|\Phi^{\lambda, R}_{J_z, P}(\xi)\rangle$.  The $N_O$ single-nucleon states are first evolved to the middle of the projection time, which are then employed to reconstruct the $N^2_S$  matrix elements.}
\label{fig:clever_sample}
\end{figure*}

The new method developed here addresses the above issue and we denote it by {\ttfamily MultSlat-Sampler\uppercase\expandafter{\romannumeral2}}. The idea is to construct a probability distribution of $\xi$ only, with $\lambda$ no longer treated as a random variable. In this way, every Monte Carlo configuration could contribute to all components of the correlation function. In order to suppress statistical fluctuation, the probability distribution sums the absolute amplitude of different Slater determinants:
\begin{equation}
\label{prob_new}
    P'(\xi) = \frac{1}{Z'}e^{-\xi^2/2}\sum_{\lambda}|\langle\Psi^{\lambda,L}_{J_z, P}(\xi)|\Psi^{\lambda,R}_{J_z, P}(\xi)\rangle|,
\end{equation}
where $Z'$ is again the normalization factor.
The price of this method is that for each configuration $\xi_k$, one needs to explicitly calculate the propagation of the $N_S$ Slater determinants, $|\Phi^{\lambda,R}_{J_z,P}(\xi_k)\rangle$ and $\langle \Phi^{\lambda,L}_{J_z,P}(\xi_k)|$, in Eq.~(\ref{Slater_evolve}), as well as the matrix element $\langle \Phi^{\lambda,L}_{J_z,P}(\xi_k)|O|\Phi^{\lambda,R}_{J_z,P}(\xi_k)\rangle$. If one performs a brute-force calculation, as shown in Figure~\ref{fig:direct_sample}, the computational cost versus $N_S$ would scale as $\mathcal{O}(N_S)$ for the former and $\mathcal{O}(N_S^2)$ for the latter. To save computational resources, we notice that the $N_S$ Slater determinants share many common orbits, so the total number of orbits, $N_O$, required to build the Slater determinants is much smaller than $N_S\times A$ for an $A$-body system. Based on this observation, as illustrated in Figure~\ref{fig:clever_sample}, we first propagate the $N_O$ orbits $\{|\phi_1\rangle,\cdots |\phi_{N_O}\rangle\}$ to the middle  of the projection time, generating $\{\langle\phi^L_1(\xi)|,\cdots \langle \phi^L_{N_O}(\xi)|\}$ and $\{|\phi^R_1(\xi)\rangle,\cdots |\phi^R_{N_O}(\xi)\rangle\}$ according to Eq.~(\ref{wedge}), then we reconstruct the matrix element using the one-body densitiy, eg, $\langle\phi^L_{k}(\xi)|\rho(\bm{n})|\phi^R_{k'}(\xi)\rangle$, given that the interpolating operator $O$ has the following form,
\begin{eqnarray}  O=\sum_{\bm{n}_1,\bm{n}_2}f(\bm{n_1-\bm{n}_2}):\rho(\bm{n_1})\rho(\bm{n}_2):.
\end{eqnarray}
As a result, the cost  is reduced from $\mathcal{O}(N_S A)$
to $\mathcal{O}(N_O)$ for propagation, and from $\mathcal{O}(N_S^2 A^2)$ to $\mathcal{O}(N_O^2)$ for measurement.

\begin{figure}
\includegraphics[height=2.4in]{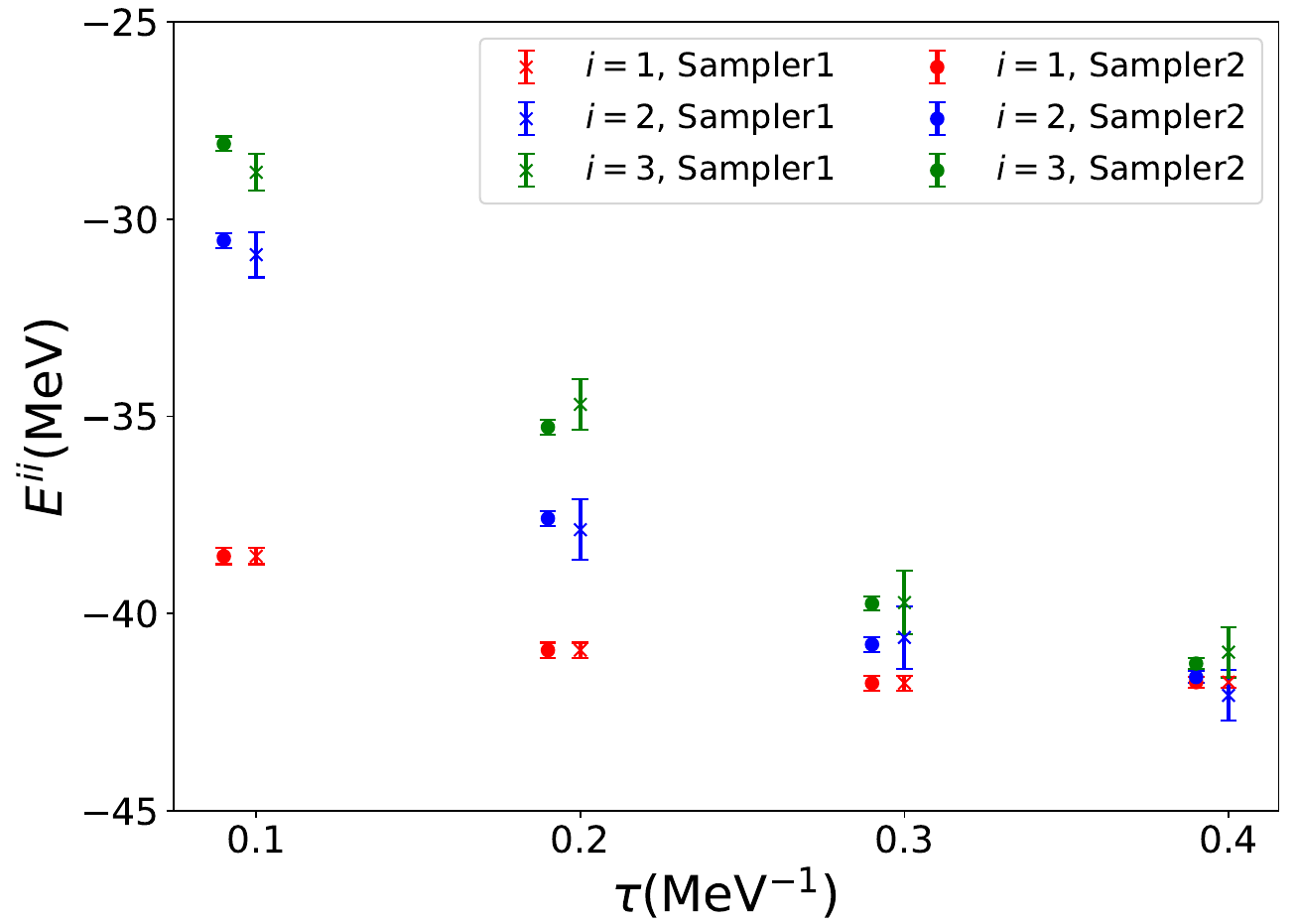}
\caption{The ratio $E^{ii}(\tau)=H^{ii}(\tau)/C^{ii}(\tau)$ of $^{7}$Li(3/2$^-$) versus the projection time $\tau$. The red, blue and green points denote the result of the three different trial states $d_1$, $d_2$ and $d_3$. The circles and the crosses represent the two methods, {\ttfamily MultSlat-Sampler\uppercase\expandafter{\romannumeral2}} and {\ttfamily MultSlat-Sampler\uppercase\expandafter{\romannumeral1}}, respectively. The error bar stands for statistical errors.}
\label{fig:algorithm_compare_woGEVP}
\end{figure}

\begin{figure}
\includegraphics[height=2.4in]{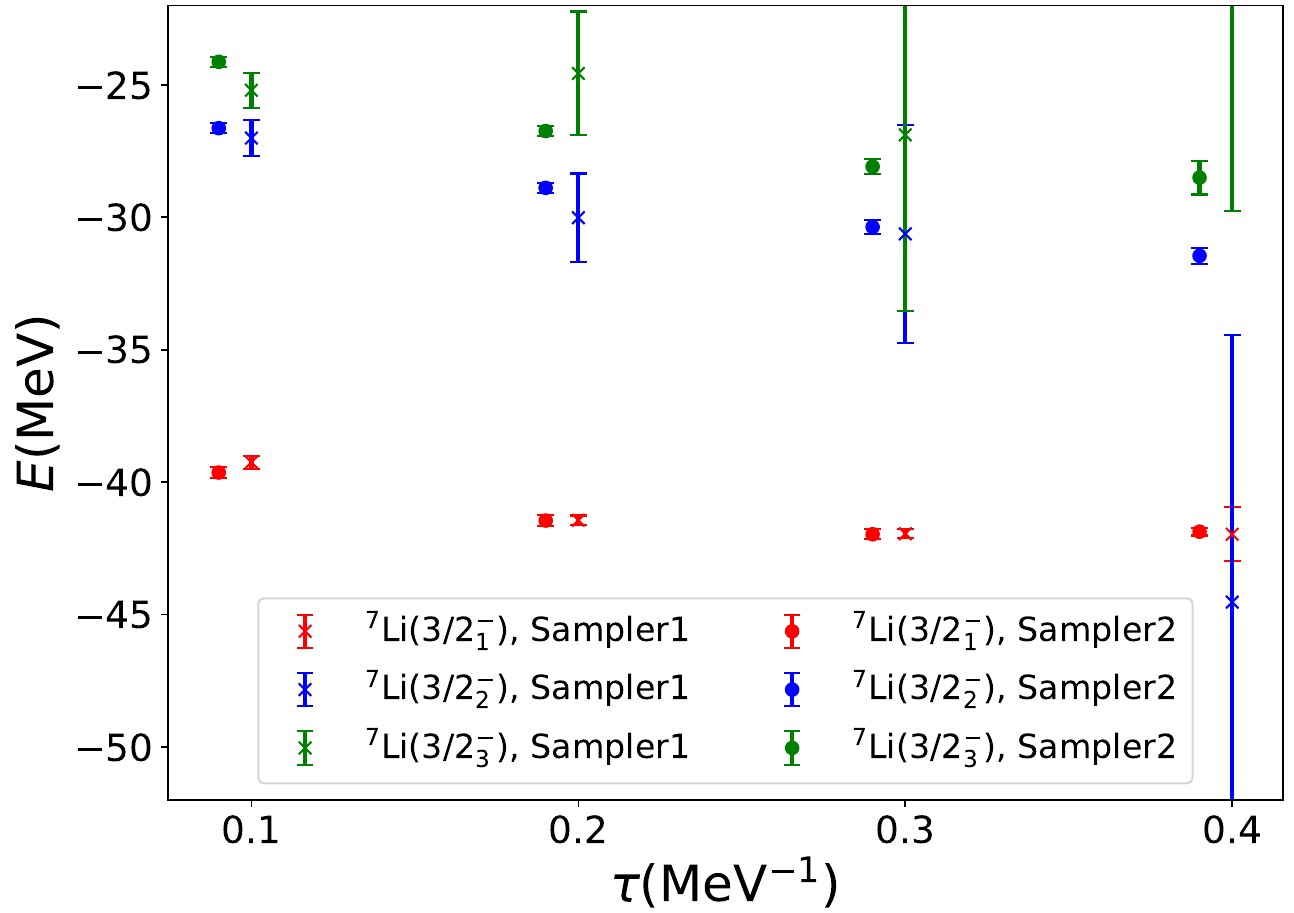}
\caption{The energy of $^{7}$Li(3/2$^-_1$), $^{7}$Li(3/2$^-_2$) and $^{7}$Li(3/2$^-_3$) versus the projection time $\tau$, which are represented by the red, blue and green points. The circles and the crosses represent {\ttfamily MultSlat-Sampler\uppercase\expandafter{\romannumeral2}} and {\ttfamily MultSlat-Sampler\uppercase\expandafter{\romannumeral1}}, respectively. }
\label{fig:algorithm_compare_wGEVP}
\end{figure}

To show the advantage of the new method, we focused on $^7$Li(3/2$^-$)  and calculated its correlation functions $H^{ij}(\tau)$  and $C^{ij}(\tau)$ using both  {\ttfamily MultSlat-Sampler\uppercase\expandafter{\romannumeral1}} and  {\ttfamily MultSlat-Sampler\uppercase\expandafter{\romannumeral2}}, from which we can analyze the energy and make a comparison. In Figure~\ref{fig:algorithm_compare_woGEVP}, we show the evolution of the ratio $E^{ii}(\tau)=H^{ii}(\tau)/C^{ii}(\tau)$ versus the projection time $\tau$, with $i=1, 2$ and 3 corresponding to the three trial states $d_1$, $d_2$ and $d_3$ in Figure~\ref{fig:Li7_conf}. Note that we have employed the same number of statistics for both methods, and the computational time required by  {\ttfamily MultSlat-Sampler\uppercase\expandafter{\romannumeral2}} is only 1.2$\sim$1.5 times more than  {\ttfamily MultSlat-Sampler\uppercase\expandafter{\romannumeral1}}. As a result, the performance of the methods can be directly judged by the magnitude of the statistical fluctuation. From Figure~\ref{fig:algorithm_compare_woGEVP}, it can be seen that the results of the two methods are consistent with each other within error bars. The statistical error is under good control for all three trial states for {\ttfamily MultSlat-Sampler\uppercase\expandafter{\romannumeral2}},  while the result of trial states $d_2$ and $d_3$ is much noisier when using {\ttfamily MultSlat-Sampler\uppercase\expandafter{\romannumeral1}}. By solving  Eq.~(\ref{GEVP_equation}), we also calculate the energies of $^{7}$Li(3/2$^-$)'s ground state and low-lying excited states. The result is presented in Figure~\ref{fig:algorithm_compare_wGEVP}. For {\ttfamily MultSlat-Sampler\uppercase\expandafter{\romannumeral2}}, the signal of both the ground state and excited states is good. In comparison, the excited states suffer from very strong statistical noise for {\ttfamily MultSlat-Sampler\uppercase\expandafter{\romannumeral1}}, and the uncertainty of the ground state increases drastically at $\tau=0.4$MeV$^{-1}$. The reason can be inferred from Figure~\ref{fig:algorithm_compare_woGEVP}, where the energies of the three trial states get very close to each other at $\tau=0.4$MeV$^{-1}$, so the large error of $d_2$ and $d_3$ pollutes the signal of the ground state in the variational calculation. In conclusion, the new algorithm outperforms the older one due to the substantial improvement of sampling efficiency,  which provides a powerful tool for accurate multi-channel calculations in NLEFT.

\section{Result}
\label{sec:Res}

In this section, we show the result of trial state improvement using $^7$Li and $^{8}$Li as an example. 

\begin{figure*}
\centering
\includegraphics[height=4.2in]{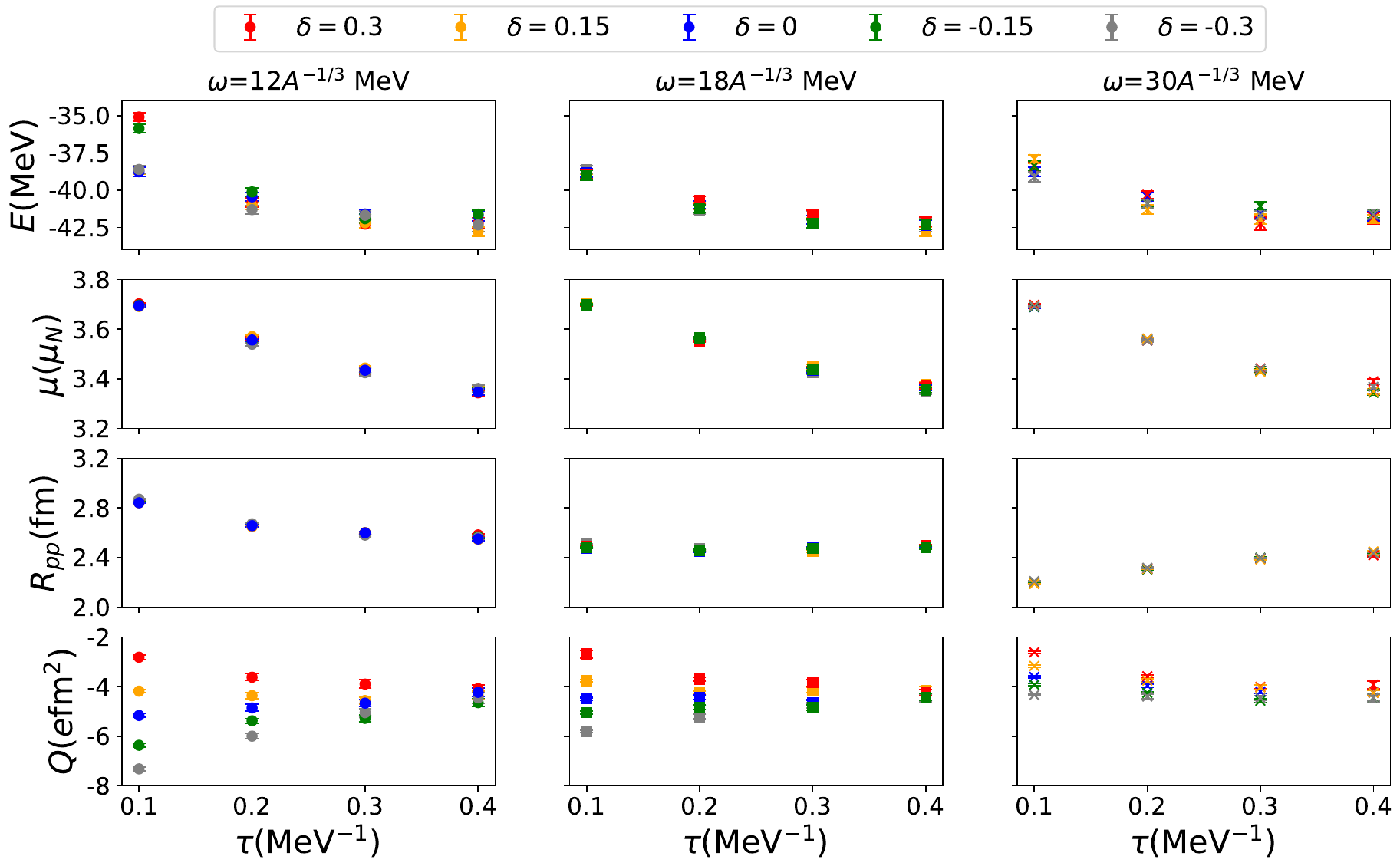}
\caption{The evolution of the energy $E$, magnetic dipole moment $\mu$, point-point charge radius $R_{pp}$ and quadrupole moment $Q$ of $^7$Li(3/2$^-_1$) versus the projection time $\tau$. The three columns from left to right represent the result of $\omega=12A^{-1/3}, 18A^{-1/3}$ and $30A^{-1/3}$ MeV in order. The red, yellow, blue, green and gray points correspond to $\delta=0.3$, 0.15, 0, -0.15, -0.3, respectively.}
\label{fig:omega_delta_opt_Li7}
\end{figure*}

We first show the result of Step 1 for $^7$Li. We used three values of the oscillator frequency, $\omega=12A^{-1/3}, 18A^{-1/3}$ and 30$A^{-1/3}$ MeV, and five values of the deformation parameter, $\delta=0.3, 0.15, 0, -0.15$ and -0.3. We calculated the energy $E$, magnetic dipole moment $\mu$, point-point charge radius $R_{pp}$ and quadrupole moment $Q$ of the ground state for each combination of $\omega$ and $\delta$, and the result is presented in Figure~\ref{fig:omega_delta_opt_Li7}. For $E$ and $\mu$, the result shows a weak dependence on $\omega$ and $\delta$. For $R_{pp}$, its evolution exhibits a strong dependence on $\omega$. If $\omega$ is too large,  the trial state would be too compact compared to the ground state, so $R_{pp}$ approaches the plateau from above as $\tau$ evolves, as is the case for $\omega=30A^{-1/3}$ MeV. In comparison, if $\omega$ is quite small ($\omega=12A^{-1/3}$MeV), the trial state would be too loose and $R_{pp}$ approaches the plateau from below. If the value of $\omega$  is appropriate, such as $18A^{-1/3}$ MeV, $R_{pp}$ would be close to the physical value at the beginning and the convergence is very fast. Note that this is consistent with previous no-core-shell-model calculations, which found that the convergence rate of the radius is sensitive to $\omega$~\cite{Choudhary:2020, Choudhary:2023}. Finally, for $Q$, both $\omega$ and $\delta$ have an influence on the evolution. Based on the above discussions, in following calculations of $^7$Li, we take $\omega^*=18A^{-1/3}$ MeV, which works the best for $R_{pp}$  and $\delta^*=0$, which converges the fastest for $Q$.

\begin{table}[htbp]
    \renewcommand\arraystretch{1.15}
	\resizebox{240pt}{!}{\scriptsize
		\begin{tabular}{c c c c }
			\hline\hline
			       & $|\Psi^{\mathrm{opt}}\rangle$         &$|\Psi^{\mathrm{compare}}\rangle$           &      EXP\\
			\hline
			   $E$(MeV)   & -42.22(14)&   -42.11(14)   & -39.24 \\
                   $\chi^2$/dof&
                1.53 & 0.92 & -\\
            \hline
               $\mu$($\mu_N$)   & 3.234(4)&   3.169(60)   & 3.256 \\
                   $\chi^2$/dof&
                1.42 & 0.78 & -\\
            \hline
               $R_{pp}$(fm)   & 2.474(3)&   2.469(9) & 2.31(5) \\
                   $\chi^2$/dof&
                 0.80& 1.25 & -\\
            \hline
               $Q$($e$fm$^2$)   & -4.32(4)&   -4.25(10) & -4.00(3) \\
                   $\chi^2$/dof&
                 1.07& 0.31 & -\\
            \hline
               $|M_1|$($\mu_N$)   & 2.90(1)&   3.04(19) & 3.16(7) \\
                   $\chi^2$/dof&
                 0.82& 0.68 & -\\
             \hline
               $|E2|$($e$fm$^2$)   & 6.91(4)&   6.81(11) & 5.78(17) \\
                   $\chi^2$/dof&
                1.07& 1.20 & - \\  
		\hline\hline
		\end{tabular}
		}
		\caption{Fitting results for  observables of $^7$Li in Figure~\ref{fig:opt_Li7}, together with the experimental data~\cite{WangMeng:2020, Nortershauser:2011, Raghavan:1989, Tilley:2002}. The value in the bracket represents the statistical error.}
	\label{tab:table1}
\end{table} 

\begin{table}[htbp]
    \renewcommand\arraystretch{1.15}
	\resizebox{240pt}{!}{\scriptsize
		\begin{tabular}{c c c c }
			\hline\hline
			       & $|\Psi^{\mathrm{opt}}\rangle$         &$|\Psi^{\mathrm{compare}}\rangle$           &      EXP\\
			\hline
			   $E$(MeV)   & -43.92(8)&   -43.94(8)   & -41.28 \\
                   $\chi^2$/dof&
                2.01 & 0.59 & -\\
            \hline
               $\mu$($\mu_N$)   & 1.444(4)&   -   & 1.654\\
                   $\chi^2$/dof&
                1.52 & - & -\\
            \hline
               $R_{pp}$(fm)   & 2.448(2)&   2.436(11) & 2.20(5) \\
                   $\chi^2$/dof&
                 1.28& 1.85 & -\\
            \hline
               $Q$($e$fm$^2$)   & 2.824(8)&   - & 3.14(2) \\
                   $\chi^2$/dof&
                 1.05& -& -\\
            \hline
               $|M_1|$($\mu_N$)   & 4.51(1)&   - & 4.1(1.4) \\
                   $\chi^2$/dof&
                 1.81& - & -\\
             \hline
               $|E2|$($e$fm$^2$)   & 4.53(2)&   4.51(6) & - \\
                   $\chi^2$/dof&
                1.04& 0.52 & - \\  
		\hline\hline
		\end{tabular}
		}
		\caption{Fitting results for  observables of $^8$Li in Figure~\ref{fig:opt_Li8}, together with the experimental data~\cite{WangMeng:2020, Nortershauser:2011, Raghavan:1989, Tilley:2004}. The value in the bracket represents the statistical error. $\mu$, $Q$ and $|M_1|$ for $|\Psi^{\mathrm{compare}}\rangle$ are not fitted  due to the poor convergence of the data.}
	\label{tab:table2}
	\end{table}

\begin{figure*}
\centering
\includegraphics[height=3.6in]{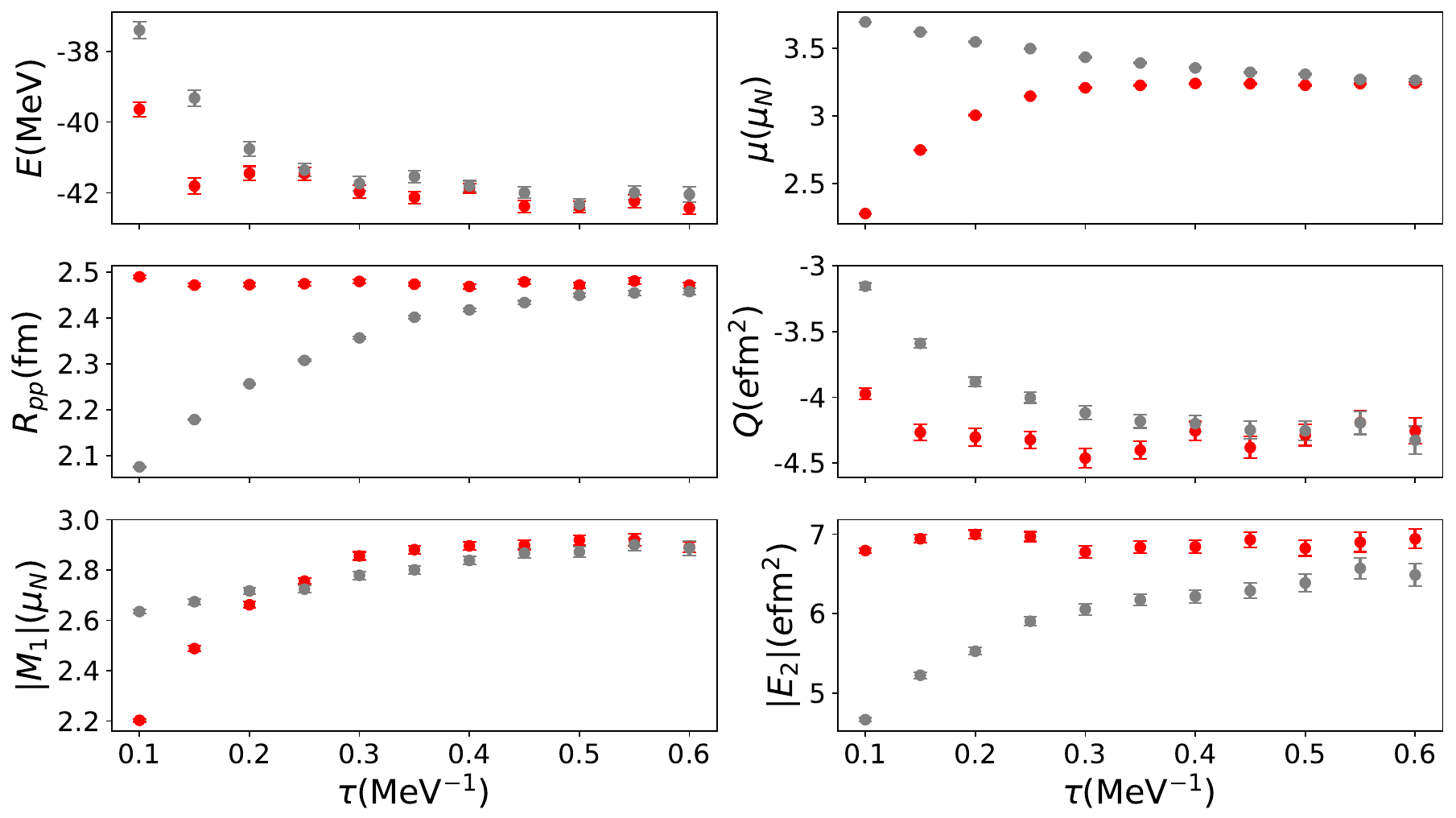}
\caption{The evolution of the energy $E$, magnetic dipole moment $\mu$, point-point charge radius $R_{pp}$ and quadrupole moment $Q$ of $^7$Li(3/2$^-_1$), as well as the $M_1$ and $E_2$ transition
matrix elements of  $^7\mathrm{Li}(1/2^-)\rightarrow ^7 \mathrm{Li}(3/2^-)$ versus the projection time $\tau$. The red and gray points represent the data of $|\Psi^{\mathrm{opt}}\rangle$ and $|\Psi^{\mathrm{compare}}\rangle$, respectively.}
\label{fig:opt_Li7}
\end{figure*}

The result of the two trial states,  $|\Psi^{\mathrm{opt}}\rangle$ and $|\Psi^{\mathrm{compare}}\rangle$, of Step 2 is presented in Figure~\ref{fig:opt_Li7}. In addition to the four observables in Step 1, we also calculated the reduced electromagnetic transition matrix elements of the $M_1$ and $E_2$ processes for $^7\mathrm{Li}(1/2^-)\longrightarrow ^7\mathrm{Li(3/2^-)}$, and one can find systematic improvement for all observables. For $E$, the improvement is the most significant when $\tau\le0.15$ MeV$^{-1}$, where the value of $|\Psi^{\mathrm{opt}}\rangle$ is significantly lower than $|\Psi^{\mathrm{compare}}\rangle$. For $\mu$, since the incorporation of $d_2$ and $d_3$ changes the valence structure of the trial state, the data points of $|\Psi^{\mathrm{opt}}\rangle$ approach the plateau from a different direction and stabilizes earlier compared to $|\Psi^{\mathrm{compare}}\rangle$. It can be expected that the convergence of $\mu$ would be further accelerated by including more valence distributions. For $R_{\mathrm{pp}}$, as explained above, the optimized oscillator frequency leads to a great acceleration of convergence.  For $Q$ and the $E_2$ matrix element, the improvement is also remarkable and the plateau begins very early. Finally, for the $M_1$ matrix element, although the deviation from the plateau of $|\Psi^{\mathrm{opt}}\rangle$ is large when $\tau$ is small, it approaches the plateau faster and outperforms $|\Psi^{\mathrm{compare}}\rangle$ at $\tau=0.3$ MeV$^{-1}$.

The fitting result of Figure~\ref{fig:opt_Li7} is given in Table~\ref{tab:table1}.  For all six observables calculated from $|\Psi^{\mathrm{opt}}\rangle$, we performed a constant fit due to the goodness of the data. For $|\Psi^{\mathrm{compare}}\rangle$, we made a constant fit to $E$ and a single-exponential fit to the rest. The results of the two different trial states are consistent within one standard error, but the error bar of $|\Psi^{\mathrm{opt}}\rangle$ is overall much smaller than $|\Psi^{\mathrm{compare}}\rangle$. Concerning $\chi^2$ per degree of freedom, it is close to one for $|\Psi^{\mathrm{opt}}\rangle$ but is too small for the $|M_1|$ matrix element calculated from $|\Psi^{\mathrm{compare}}\rangle$, indicating a possible overestimation of the uncertainty.  In Table~\ref{tab:table1}, we also list the experimental data based on which an investigation of systematic effects neglected in this work can be carried out. Compared to the experiment, the overall agreement is acceptable considering the simple form of the interaction we have used. However, the lattice result is a bit too loose for $R_{pp}$ and  too bound for $E$, and the deformation is a bit too large concerning $Q$ and $|E_2|$. These deviations might be traced back to the absence of certain important components of the interaction, such as the tensor force and other three-body forces. In addition, as nuclear electric
quadrupole observables are sensitive to large-distance tails of the nuclear wave
function~\cite{Caprio:2022}, further investigations of finite-volume dependence of $Q$ and $|E_2|$ are essential in future works.

\begin{figure*}
\centering
\includegraphics[height=3.6in]{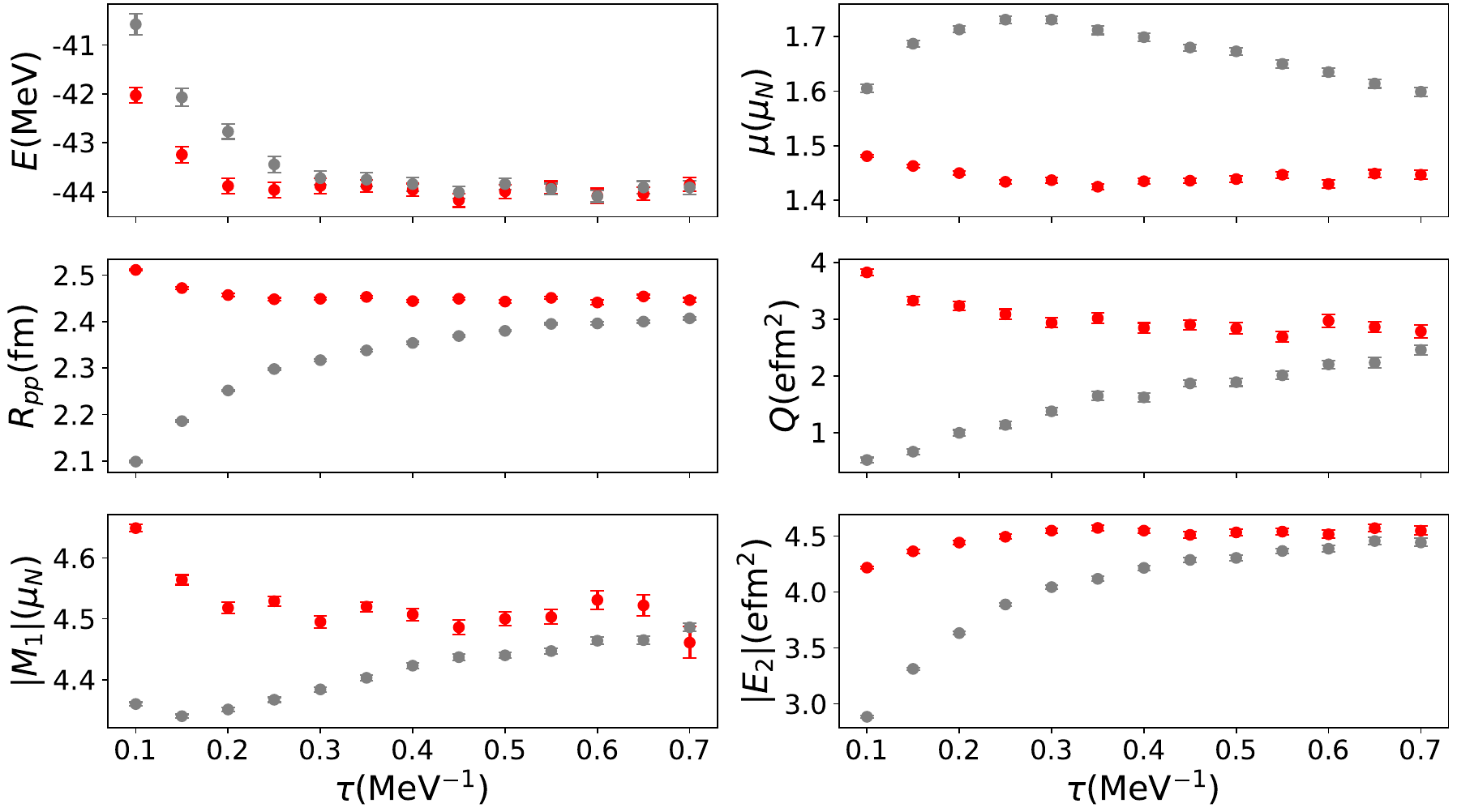}
\caption{The evolution of the energy $E$, magnetic dipole moment $\mu$, point-point charge radius $R_{pp}$ and quadrupole moment $Q$ of $^8$Li(2$^+_1$), as well as the $M_1$ and $E_2$ transition
matrix elements of  $^8\mathrm{Li}(1^+)\rightarrow ^8 \mathrm{Li}(2^+)$ versus the projection time $\tau$. The red and gray points represent the data of $|\Psi^{\mathrm{opt}}\rangle$ and $|\Psi^{\mathrm{compare}}\rangle$, respectively.}
\label{fig:opt_Li8}
\end{figure*}

For $^8$Li, we followed the same procedure of $^7$Li and determined the optimized values of the oscillator frequency and deformation parameter as $\omega^*=18A^{-1/3}$ MeV and $\delta^*=0.15$, and the comparison between $|\Psi^{\mathrm{opt}}\rangle$ and $|\Psi^{\mathrm{compare}}\rangle$ is shown in Figure~\ref{fig:opt_Li8}. Again, the plateau of $|\Psi^{\mathrm{opt}}\rangle$ begins quite early, while no observable except the energy reaches the plateau up to  $\tau=0.7$ MeV$^{-1}$ for $|\Psi^{\mathrm{compare}}\rangle$. Note that compared to $^7$Li, the improvement of $^8$Li is more significant in two aspects. Firstly, different from the energy, the evolution of certain electromagnetic observables is not necessarily monotonic, especially for $\mu$ calculated from $|\Psi^{\mathrm{compare}}\rangle$, which reflects the redistribution of valence nucleons under the operation of the imaginary-time projector. This non-monotonic behavior may greatly complicate the fit and even cause a fake plateau when large fluctuations exist. By including various valence excitations in the trial state $|\Psi^{\mathrm{opt}}\rangle$, the evolution becomes monotonic again. Secondly, the convergence of $^8$Li without trial state improvement is not as good as that of $^{7}$Li at large $\tau$. For example, the gray points for $\mu$, $Q$ and $|M_1|$ in Figure~\ref{fig:opt_Li8} are almost linear versus $\tau$ in the asymptotic region, which do not allow a stable multi-exponential fit. After optimizing the trial state, there is a clear plateau and one can employ a simple constant fit. The non-monotical behavior and slow convergence rate observed in $|\Psi^{\mathrm{compare}}\rangle$ is due to the mixing of ground state and excited state wave functions, see Appendix~\ref{app:appendixB}  for a detailed explanation. We also refer to  the supplemental material of Ref.~\cite{Lu:2021} for relevant discussions. The fitting result of Figure~\ref{fig:opt_Li8} is shown in Table~\ref{tab:table2}. $\mu$, $Q$ and $|M_1|$ of $|\Psi^{\mathrm{compare}}\rangle$ are not fitted due to the above reason. The necessity of our method for imaginary time extrapolation is well demonstrated by Table~\ref{tab:table2}. Without trial state improvement, the extrapolation would be extremely challenging and even impossible.

 \section{Summary}
\label{summary}

Nuclear Lattice Effective Field Theory (NLEFT) is a powerful $ab$ $initio$ framework for addressing nuclear many-body problems. Like other $ab$ $initio$ methods, however, it is subject to systematic uncertainties that constrain its predictive accuracy. In this work, we target one prominent source of uncertainty: the error associated with imaginary-time extrapolation, a critical component of NLEFT calculations~\cite{Lahde:2014,Lahde:2015}. To mitigate this issue, we develop an approach for implementing multi-reference trial wave functions in lattice Quantum Monte Carlo simulations that more closely approximate the true ground state from the outset. We demonstrate the method on $^7$Li and $^8$Li, achieving accelerated convergence and substantially reduced extrapolation uncertainties across a range of observables. The improvement is especially pronounced for electromagnetic observables, which are highly sensitive to nuclear structure and otherwise exhibit prohibitively slow convergence without optimized trial states. We further benchmark NLEFT predictions against experimental data to assess the interaction developed in~\cite{Niu:2025}, finding reasonable agreement overall while highlighting the need to incorporate missing components of the nuclear force.

A key technical challenge in realizing multi-reference trial states within NLEFT is the efficient computation of correlation functions between different Slater determinants sampled from a common Monte Carlo ensemble. Here, we introduce an efficient sampling algorithm that resolves this difficulty. Comparisons with previous methods reveal significant enhancements in signal quality for both ground and excited states. This advance provides a robust foundation for multi-reference trial states and opens promising pathways towards accurate multi-channel calculations of nuclear spectra, properties, and transitions in future NLEFT studies.

The multi-reference trial states considered here are constructed from shell-model single-particle wave functions, making them particularly suitable for nuclei exhibiting pronounced shell structure. For states with exotic shapes or clustering, such bases may yield limited overlap. Nonetheless, our framework offers a general procedure for building optimized multi-reference trial states even in such cases. A prominent example is $\alpha$-cluster structures~\cite{Oertzen:2006, RevModPhys.90.035004, Horiuchi:2012}, prevalent in light nuclei—such as the Hoyle state in $^{12}$C~\cite{Hoyle:1954}—and at the surfaces of heavier nuclei where $\alpha$ formation and decay occur. Cluster configurations of this type have already been explored in NLEFT calculations~\cite{Shen:2023}. It would therefore be interesting in future work to combine such cluster configurations with the present multi-reference optimization strategy, which may further improve the description of nuclear clustering phenomena. Considering the prevalence of clustering and mean-field duality in light nuclei~\cite{RevModPhys.90.035004, Shen:2023, Arumugam:2005}, the multi-reference method introduced here could naturally take into account  both sides of effects,  which would hopefully improve the accurate calculation of  structures and reactions of light nuclei.

\begin{acknowledgments}

X.F. and T.W. were supported in part by NSFC of China under Grants No. 12125501 and No. 12550007. B.N.L. was supported by NSAF No. U2330401 and National Natural Science Foundation
of China with Grant Nos. 12275259, 12547105.

\end{acknowledgments}

\newpage
\bibliography{trialstate}
\newpage

\begin{widetext}
\appendix
\setcounter{page}{1}
\renewcommand{\thepage}{Supplementary Information -- S\arabic{page}}
\setcounter{table}{0}

\section{Shell Distributions for the Trial States of $^7$Li(1/2$^-$), $^8$Li(2$^+$) and $^8$Li(1$^+$)}

In Figure~\ref{fig:Li7_1ov2_conf}, \ref{fig:Li8_2_conf} and \ref{fig:Li8_1_conf}, we show the shell distributions used in this work to construct the trial states of $^7$Li(1/2$^-$), $^8$Li(2$^+$) and $^8$Li(1$^+$), respectively. 

\begin{figure}[htbp]
\includegraphics[height=1.8in]{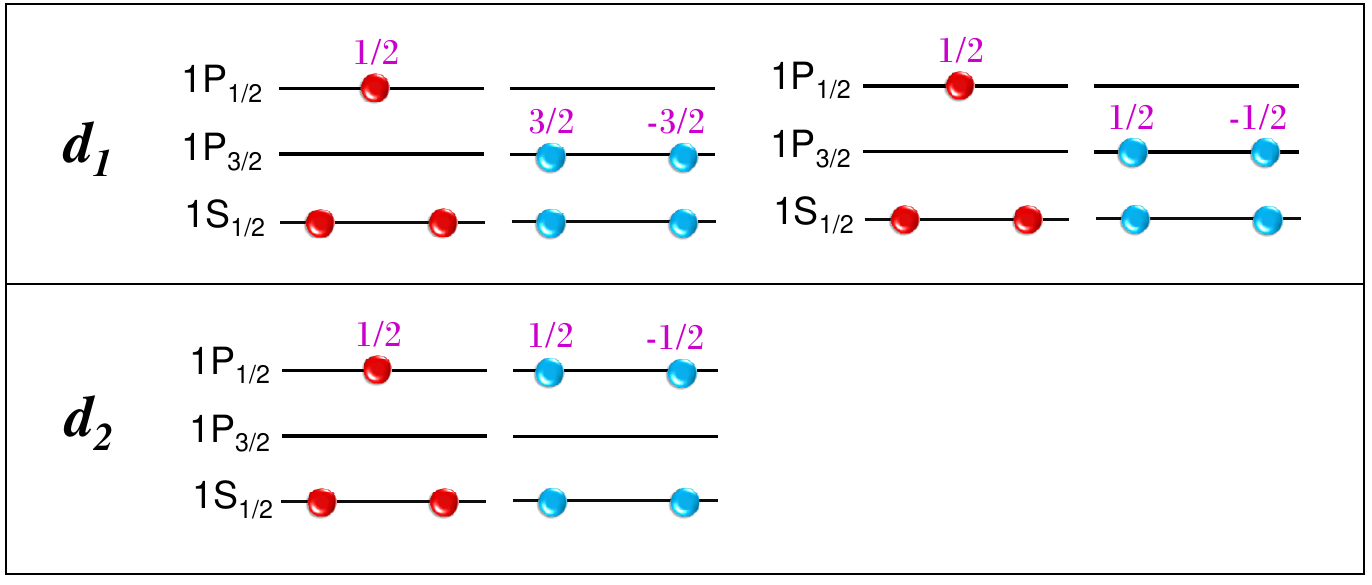}
\caption{The two different shell distributions, $d_1$ and $d_2$, for constructing the trial state of $^7$Li(1/2$^-$) in this work. The meaning of the symbols here is the same as Figure~\ref{fig:Li7_conf}.}
\label{fig:Li7_1ov2_conf}
\end{figure}

\begin{figure}[htbp]
\includegraphics[height=3.2in]{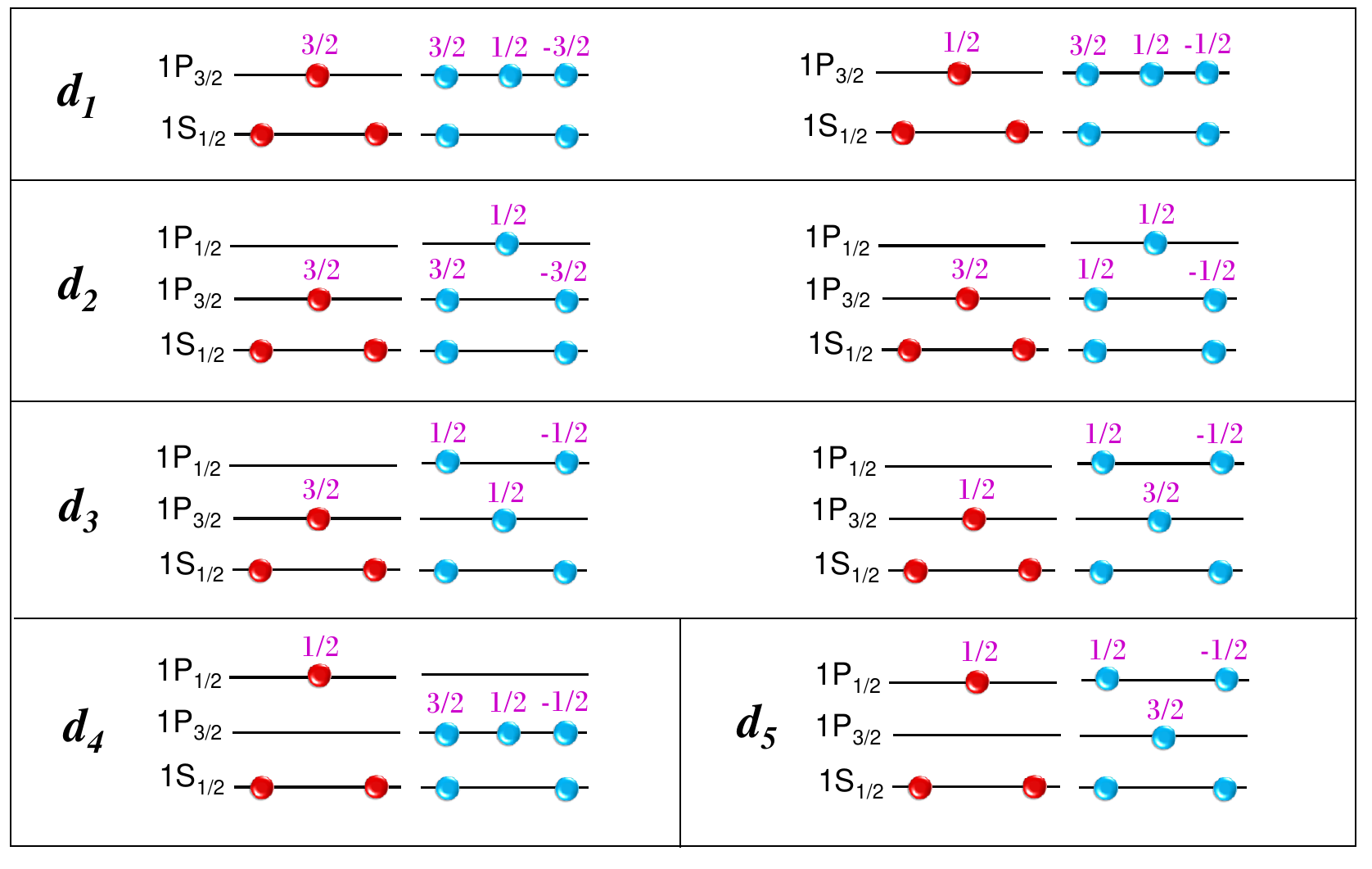}
\caption{The five different shell distributions, $d_1$, $d_2$, $d_3$, $d_4$ and $d_5$, for constructing the trial state of $^8$Li(2$^+$) in this work. The meaning of the symbols here is the same as Figure~\ref{fig:Li7_conf}.}
\label{fig:Li8_2_conf}
\end{figure}

\begin{figure}[htbp]
\includegraphics[height=3.0in]{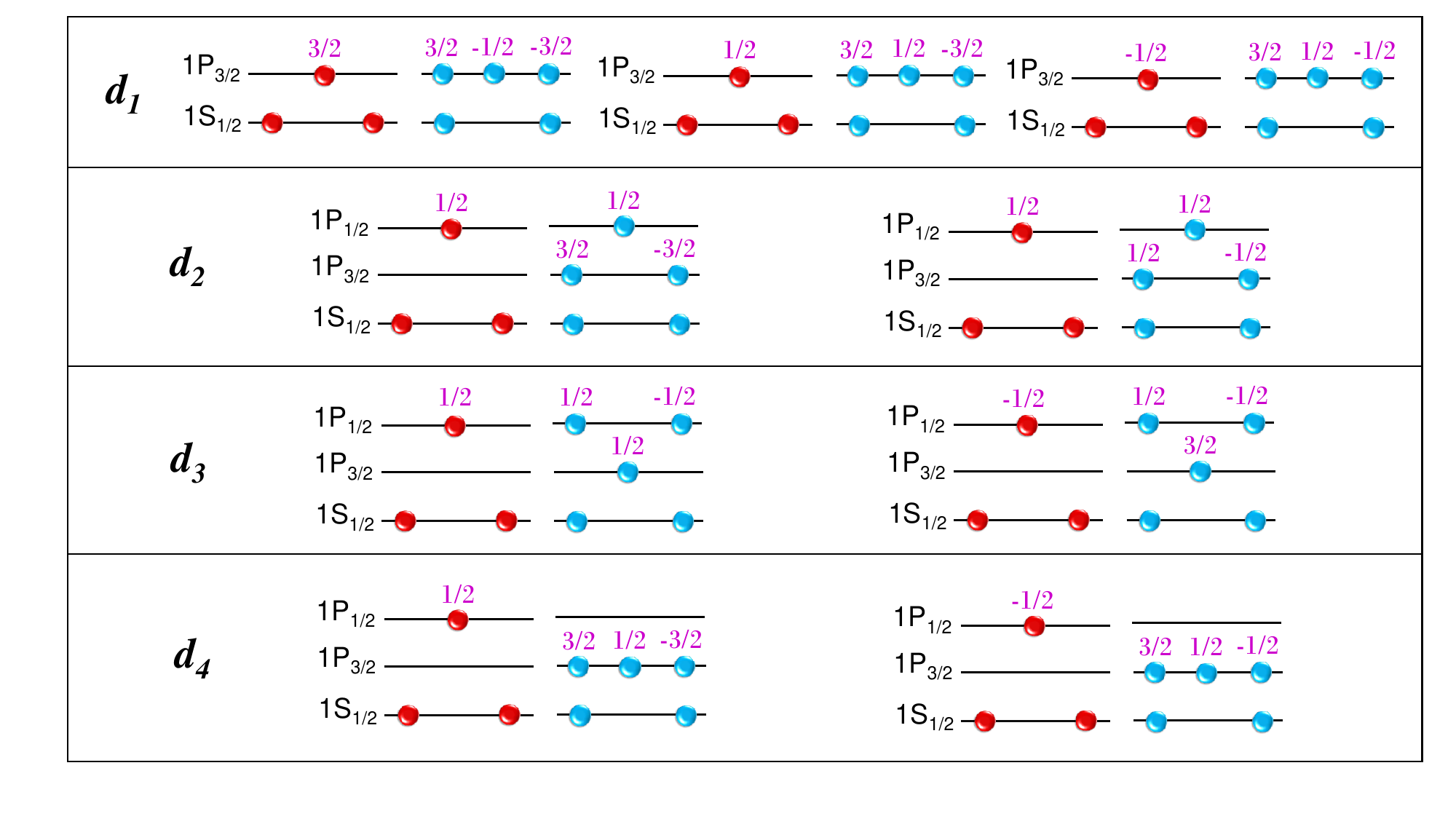}
\caption{The four different shell distributions, $d_1$, $d_2$, $d_3$ and $d_4$, for constructing the trial state of $^8$Li(1$^+$) in this work. The meaning of the symbols here is the same as Figure~\ref{fig:Li7_conf}. }
\label{fig:Li8_1_conf}
\end{figure}

\newpage
\section{A Model for Understanding the Imaginary Time Evolution of Nuclear Electromagnetic Observables}
\label{app:appendixB}

In Figure~\ref{fig:opt_Li8}, we find that for  electromagnetic observables of $^{8}$Li, the convergence of imaginary time projection is generally slow using the unoptimized trial state $|\Psi^{\mathrm{compare}}\rangle$. The situation is even more strange for the magnetic moment $\mu$, which exhibits a non-monotonic dependence on $\tau$. To illustrate the origin, we consider three lowest-lying eigenstates of $^{8}$Li($2^+$) for the lattice Hamiltonian $H$ employed in this work, each labeled as $|\Phi_0\rangle$, $|\Phi_1\rangle$ and $|\Phi_2\rangle$ from low to high. The trial state $|\Psi^{\mathrm{compare}}\rangle$ is modeled as
\begin{equation}
\label{eq:Psi_model}    |\Psi^{\mathrm{compare}}\rangle = a_0|\Phi_0\rangle+a_1|\Phi_1\rangle+ a_2|\Phi_2\rangle,
\end{equation}
where contributions of higher excited states are neglected. Operating the imaginary time projector $e^{-H\tau/2}$ on $|\Psi^{\mathrm{compare}}\rangle$, the system is evolved into
\begin{equation}
  |\Psi(\tau)\rangle\equiv e^{-H\tau/2}|\Psi^{\mathrm{compare}}\rangle=\sum_{k=0}^2 a_ke^{-
  (E_0+\Delta E_k)\tau/2}|\Phi_k\rangle,
\end{equation}
where $E_0$ is the ground-state energy, while $\Delta E_1$ and $\Delta E_2$ are excitation energies of the first and second excited states, respectiveply. $\Delta E_0 = 0$ is introduced for convenience. The expectation value of an observables $O$ with respect to $|\Psi(\tau)\rangle$ is given by,
\begin{equation}
\label{eq:O_tau}
    O(\tau)\equiv \frac{\langle \Psi(\tau)|O|\Psi(\tau)\rangle}{\langle \Psi(\tau)|\Psi(\tau)\rangle}=\frac{\sum_{k,k'=0}^2a_k a_{k'}^* e^{-(\Delta E_k+\Delta E_k')\tau/2}\langle \Phi_{k'}|O|\Psi_k\rangle}{\sum_{k=0}^2|a_k|^2  e^{-\Delta E_k\tau}}.
\end{equation}
When $O$ is taken as $H$, the non-diagonal matrix element vanishes, i.e. $\langle\Phi_{k'}|H|\Phi_k\rangle=0$ for $k\neq k'$, so the expression above is reduced to
\begin{equation}
\label{eq:E_tau}
    E(\tau) = \frac{\sum_{k=0}^2 |a_k|^2  e^{-\Delta E_k \tau} E_k}{\sum_{k=0}^2 |a_k|^2  e^{-\Delta E_k\tau}},
\end{equation}
clearly the excited-state contanimation is suppressed by the exponential $e^{-\Delta E_1\tau}$ which decays rapidly. Moreover, since all the coefficients in Eq.~(\ref{eq:E_tau}) are positive-definite, $E(\tau)$ always decreases monotonically as $\tau$ increases. The features above give rise to the fast convergence of $E(\tau)$, which can be fitted accurately. For a general operator $O$ such as electromagnetic properties, however, the non-diagonal matrix element generally survives and its value can be considerable, hence the supression factor for excited-state contanimation is $e^{-\Delta E_1\tau/2}$, which decays more slowly. Besides, the coefficents multiplying the exponential in the numerator of Eq.~(\ref{eq:O_tau}) can be either positive or negtive, which is the origin for the non-monotonic behavior we just mentioned. 

\begin{figure}[htbp]
\includegraphics[height=5.0in]{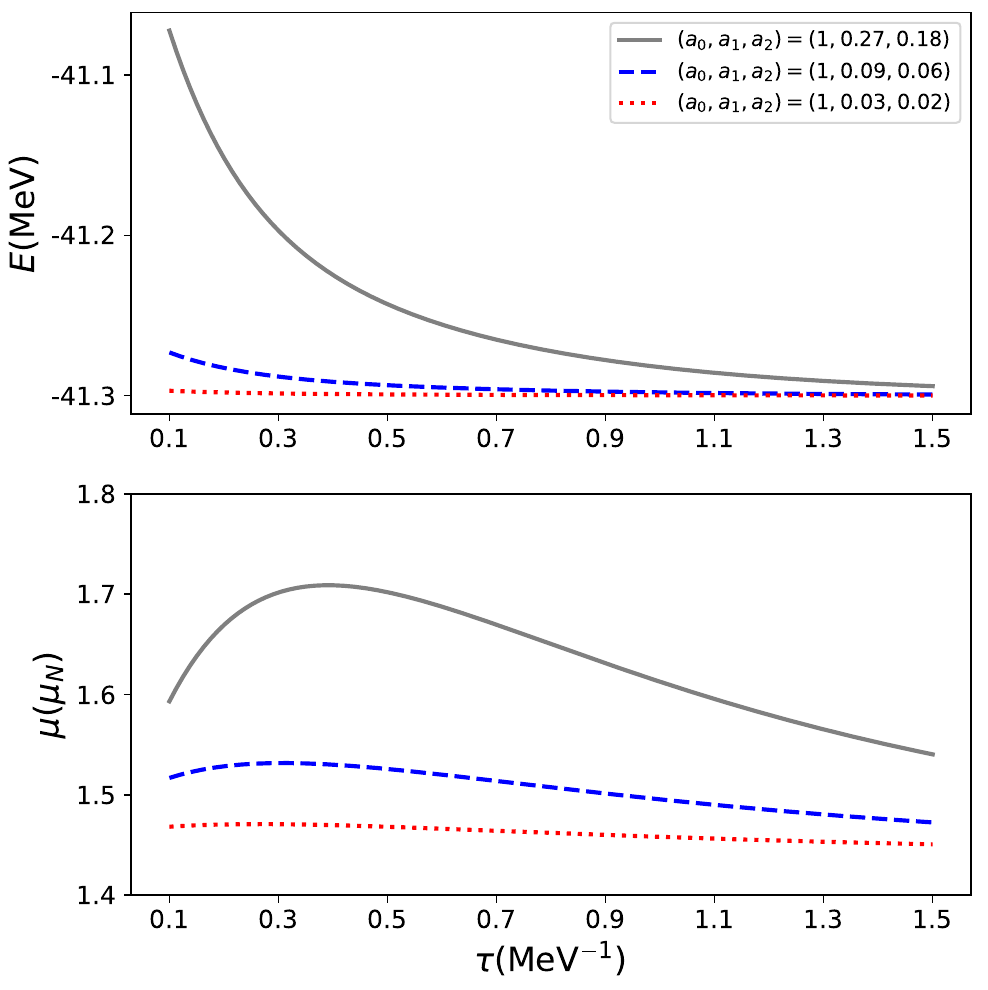}
\caption{The dependence of the energy $E$ and magnetic moment $\mu$ versus $\tau$ using different trial states in our model. The gray solid line, blue dashed line and red dotted line represent the three combinations of $(a_0, a_1, a_2)$ in Eq.~(\ref{eq:a_combination}).}
\label{fig:EM_model}
\end{figure}

To visualize the analysis above, we determine the energies $E_0$, $E_1$, $E_2$ and the matrix element of $\mu$ for $^{8}$Li($2^+$) from the lattice date, whose central values are
\begin{equation}
    E_0=-41.3 \ \mathrm{MeV},\quad E_1 = -39.1\ \mathrm{MeV},\quad E_2 = -33.5\ \mathrm{MeV}\quad \Rightarrow\quad \Delta E_1=2.2\ \mathrm{MeV},\quad \Delta E_2 = 7.8\ \mathrm{MeV},  
\end{equation}
and
\begin{equation}
    \begin{bmatrix}    \langle\Phi_0|\mu|\Phi_0\rangle, & \langle\Phi_0|\mu|\Phi_1\rangle, & \langle\Phi_0|\mu|\Phi_2\rangle\\
        \langle\Phi_1|\mu|\Phi_0\rangle, & \langle\Phi_1|\mu|\Phi_1\rangle, & \langle\Phi_1|\mu|\Phi_2\rangle\\
        \langle\Phi_2|\mu|\Phi_0\rangle, & \langle\Phi_2|\mu|\Phi_1\rangle, & \langle\Phi_2|\mu|\Phi_2\rangle
    \end{bmatrix} = \begin{bmatrix}
        1.44\mu_N,&0.94\mu_N,&-0.77\mu_N\\
        0.94\mu_N,&3.01\mu_N,&-2.66\mu_N\\
        -0.77\mu_N,&-2.66\mu_N,&-1.08\mu_N
    \end{bmatrix}. 
\end{equation}
Statistcal errors associated with the above quatities are neglected  when constructing the model. We then consider the evolution of $E(\tau)$ and $\mu(\tau)$  for the following three combinations of $(a_0, a_1, a_2)$ in Eq.~(\ref{eq:Psi_model}), 
\begin{equation}
\label{eq:a_combination}
\begin{aligned} &\mathrm{\uppercase\expandafter{\romannumeral1}}.\ (a_0, a_1,a_2) = (1, 0.27, 0.18),\\ 
   &\mathrm{\uppercase\expandafter{\romannumeral2}}. \ (a_0, a_1,a_2) = (1, 0.09, 0.06),\\
   &\mathrm{\uppercase\expandafter{\romannumeral3}}. \ (a_0, a_1,a_2) = (1, 0.03, 0.02),
\end{aligned}
\end{equation}
where $a_0=1$ is fixed and the prefactors $a_1$ and $a_2$ are successively scaled by $1/3$ from case $\mathrm{\uppercase\expandafter{\romannumeral1}}$ to case $\mathrm{\uppercase\expandafter{\romannumeral3}}$, representing the suppresion of excited-state components in $|\Psi^{\mathrm{compare}}\rangle$. Note that even for case $\mathrm{\uppercase\expandafter{\romannumeral1}}$, the ground-state proportion in $|\Psi^{\mathrm{compare}}\rangle$ is already much larger than other excited states. In the upper and lower panels of Figure~\ref{fig:EM_model}, we show the evolution of $E$ and $\mu$ versus $\tau$ respectively. For $E$, it evolves  monotonically regardless of the value of $(a_0, a_1, a_2)$. Since we have neglected the statistical error of $E$ in Figure~\ref{fig:EM_model}, which is of the order $0.1\mathrm{MeV}$, clearly the three combintations are no longer distinguishable once statistical noises are incorporated. This means that the determination of energy is not very sensitive to the trial state employed. On the contrary, the evolution of $\mu$ is more sensitive to excited-state effects. For case $\mathrm{\uppercase\expandafter{\romannumeral1}}$, convergence is not reached even up to $\tau=1.5\mathrm{MeV}^{-1}$, and its non-monotonical behavior  is in agreement with the functional form  of $\mu(\tau)$ calculated from $|\Psi^{\mathrm{compare}}\rangle$ in Figure~\ref{fig:opt_Li8}. The trial state improvement method developed in this work amounts to a systematic control of excited-state contamination, leading to much improved trial states similar to case $\mathrm{\uppercase\expandafter{\romannumeral2}}$ and case $\mathrm{\uppercase\expandafter{\romannumeral3}}$ above.

\end{widetext}

\end{document}